\documentclass[a4paper,11pt]{article}
\usepackage[utf8]{inputenc}
\usepackage{amsfonts}
\usepackage{amssymb}
\usepackage{graphicx}
\usepackage{fancyhdr}
\usepackage{fancyvrb}
\usepackage{fancybox}
\usepackage{float}
\usepackage{geometry}
\usepackage{minitoc}
 \usepackage{indentfirst}
 \usepackage{multicol}
 \usepackage[outerbars]{changebar}
 \usepackage{pifont,textcomp}
 \usepackage{amsfonts}
 \usepackage{amsmath}
 \usepackage{amscd}
 \usepackage{array}
\usepackage{multirow}
 \usepackage{mathrsfs}
 \usepackage{amssymb}
 \usepackage{amsthm}

\title{Minimal and  maximal   lengths  of quantum gravity from  non-Hermitian position-dependent  noncommutativity}
\author{ {Lat\'evi M. Lawson}\\
\space\\
 latevi@aims.edu.gh
}

\begin{document}
\maketitle

\begin{abstract}
A minimum length scale of the order of Planck length is a feature of many models of quantum gravity that seek to unify quantum mechanics and
gravitation. Recently, Perivolaropoulos in his seminal work [Phys. Rev.D 95, 103523 (2017)] predicted the simultaneous existence of minimal and maximal length measurements of  quantum gravity. More recently, we have shown that both
measurable lengths can be obtained from position-dependent noncommutativity [J. Phys. A: Math.Theor. 53, 115303 (2020)]. In this paper, we present an alternative derivation of these lengths from non-Hermitian position-dependent noncommutativity. We show that a simultaneous measurement of both lengths form a family of discrete  spaces. In one hand, we show the similarities between the maximal uncertainty measurement and the classical properties of gravity. On the other hand, the connection between the minimal uncertainties and the non-Hermicity quantum mechanic scenarios. The existence of minimal uncertainties are the consequences of non-Hermicities of some operators that are generators of this noncommutativity. With an appropriate Dyson map, we demonstrate by a similarity transformation that the physically  meaningfulness of dynamical quantum systems is generated by a hidden Hermitian position-dependent noncommutativity. This transformation preserves the properties of quantum gravity but  removes the fuzziness induced by minimal uncertainty measurements at this  scale. Finally, we study the eigenvalue problem of a free particle in a square-well potential in these new
Hermitian variables.


\end{abstract}


\section{Introduction}
The idea of noncommutativity of space-time might provide deep indications about the quantum nature of space-time at a very small distance, where a full theory of quantum gravity must be invoked, has its root in string theory \cite{3}. In
fact, the noncommutativity of space-time is one of the promising candidate theories to the unification of quantum theory and General Relativity (GR). All the
other candidate theories of unification such as string theory \cite{4}, black hole theory
\cite{5}, loop quantum gravity \cite{6} predicted the existence of minimal measurement of
quantum gravity at the Planck scale. To theoretically realize this minimal length scale in quantum mechanics, one has introduced a simple model, the so-called Generalized Uncertainty Principle (GUP) \cite{7,8,9} which is a gravitational correction
to quantum mechanics. Mostly, these theories of quantum gravity are restricted to the case where there is a nonzero minimal uncertainty in the position. Only Doubly Special Relativity (DSR) theories \cite{10,11,12} suggest an addition to the
minimal length, the existence of a maximal momentum. Recently, Perivolaropoulos proposed a consistent algebra that induces for a simultaneous measurement, a maximal length and a minimal momentum \cite{1}. In this approach, the maximal length
of quantum gravity is naturally arisen in cosmology due to the presence of particle horizons.  Perivolaropoulos also predicted the simultaneous existence of maximal and minimal position uncertainties. More recently, we have shown that both position uncertainties can simultaneously be obtained from
position-dependent noncommutativity and the minimal momentum is provided by the position-dependent deformed Heisenberg algebra \cite{2}. In continuation of this
work, we show that both lengths can also be  derived from non-Hermitian position-dependent noncommutativity. The simultaneous presence of both lengths
at this scale form a lattice system in which each site represented by the minimal length is spaced by the maximal length. At each minimal length point results of
the unification of magnetic  and gravitational fields. As has been recently shown \cite{12'}, the maximal length allows probing quantum gravitational effects with low
energies and manifests properties similar to the classical ones of General Relativity (GR).

It is well known that the existence of minimal uncertainties in quantum mechanics induces among other consequences \cite{13, 14, 15, 16, 17} a non Hermicity of some operators that generate the corresponding Hilbert space \cite{6,18,19,20}.
In the present case, the minimal length in the X-direction and the minimal momentum in the $P_y$-direction lead to the non-Hermiticity of operators $\hat X$ and $\hat P_y$
that generate the noncommutative space \cite{1}. Consequently, Hamiltonians $\hat H$ of systems involving these operators will in general also not be Hermitian. The
corresponding eigenstates no longer form an orthogonal basis and the Hilbert space structure will be modified. In order to map these operators into their Hermitian
counterparts. We introduce a positive-definite Dyson map $\eta$ \cite{21} and its associated
metric operator $\rho$ which generate a hidden Hermitian position-dependent noncommutativity by means of similarity transformation of the non-Hermitian one
i.e $\eta\left( \hat X,\hat Y,\hat P_x,\hat P_y,\hat H   \right)\eta= \left( \hat x,\hat y,\hat p_x,\hat p_y,\hat h   \right)=\left( \hat x^\dag,\hat y^\dag,\hat p_x^\dag,\hat p_y^\dag,\hat h^\dag   \right)  $. Doing so, we
tie a connection between the quantum mechanic noncommutativity with GUP \cite{22,23,24,25,26,27,28,28',28'',28'''} and the non-Hermiticity quantum mechanic scenarios
\cite{29,30,31,32,33,34,35,36,37,38,39}. Furthermore, this transformation preserves
the uncertainty measurements at this scale but  removes the fuzziness induced by the minimal uncertainty measurements. Finally, within this hidden Hermitian
space, we present the eigensystems of a free particle in a box. We show that
the existence of maximal length induces strong quantum gravitational effects in this box. These effects are manifested by the deformations of quantum energy and these deformations are more pronounced as one increases the quantum levels, allowing the particle to jump from one state to another with low energies and
with high probability densities \cite{12'}. These properties are similar to the classical gravity of General relativity
where the gravitational field becomes stronger for heavy systems that curve the
space, enabling the surrounding light systems to fall down with low energies. The
resulting time inside of this space runs out more slowly as the gravitational effects increase.

In what follows, we explore in section 2, the similarities between our recent
deformed noncommutativity with GUP and the pseudo-Hermiticity quantum
mechanic scenarios. We show that these deformations lead to a non Hermiticity of the position operator $\hat X$  and the momentum operator $\hat P_y$. By constructing a
Dyson map \cite{21} we provide their corresponding set of Hermitian counterparts.
As a consequence of these deformations, we derive in section 3, the uncertainty
measurements resulting from these deformations. In section 4, we study in terms
of our new set of variables, the model of particles in a 2D box. We present our
conclusion in section 5.

\section{ Non-Hermitian position dependent noncommutativity } \label{sec2}
Given a set  operators of $\hat X, \hat Y, \hat P_x, \hat P_y$  defined on  the 2D Hilbert space   and satisfy the following commutation relations and all possible permutations of the Jacobi identities \cite{1}
\begin{eqnarray} \label{alg1}
[\hat X,\hat Y]&=&i\theta (1-\tau\hat Y +\tau^2 \hat Y^2),\quad [\hat X,\hat P_x ]=i\hbar (1-\tau \hat Y +\tau^2 \hat Y^2),\cr
 {[\hat Y,\hat P_y ]}&=&i\hbar (1-\tau \hat Y +\tau^2 \hat Y^2),\quad
[\hat X,\hat P_y]=i\hbar\tau(2\tau \hat Y\hat X-\hat X)+
  i\theta\tau(2\tau \hat Y\hat P_y-\hat P_y)\cr
 { [\hat P_x,\hat P_y]}&=&0, \quad \quad\quad \quad\quad \quad \quad\quad \quad  {[\hat Y,\hat P_x]}=0. 
\end{eqnarray}
where $\theta,\tau \in \mathbb{R}_+^*$  are both   deformed parameters that describe  the frontier of the Planck scale.
The parameter $\tau$ is the GUP  deformed parameter \cite{7,43,44} related to  quantum gravitational effects  at this scale.
The parameter $\theta$ is related to the    noncommutativity of the space at this scale  \cite{29,47,48}. In the framework of noncommutative classical or quantum mechanics, this parameter  is proportional to the inverse of a constant magnetic   field such that  $\theta=1/B$ \cite{48,49,50}. Since the algebra (\ref{alg1}) describes the space  at the  Planck scale, then such  magnetic  fields   are  necessarily  superstrong and  may play the  role of primordial magnetic fields.
Obviously by taking $\tau\rightarrow 0$,  we get the $\theta$-deformed  space 
  \begin{eqnarray}\label{alg2}
 [\hat x_0,\hat y_0]&=&i\theta,\,\,\,[\hat x_0,\hat p_{x_0}]=i\hbar,\,\,\,[\hat y_0,\hat p_{y_0}]=i\hbar,\cr
 {[\hat p_{x_0},\hat p_{y_0}]}&=&0,\,\,\quad [\hat x_0,\hat p_{y_0}]=0,\,\,\, [\hat y_0, \hat p_{x_0}]=0.
\end{eqnarray}
Using the   asymmetrical Bopp-shift \cite{38}, we can relate the noncommutative  operators (\ref{alg2}) to   the ordinary commutations ones as follows:
 \begin{eqnarray}\label{bob}
 \hat x_0=\hat x_s-\frac{\theta}{2\hbar}\hat  p_{y_s},\quad
 \hat y_0=\hat y_s,
\end{eqnarray}
where the Hermitian operators $\hat x_s,\hat y_s,\hat p_{x_s},\hat  p_{y_s}$ satisfy the ordinary $2$D Heisenberg algebra
\begin{eqnarray}\label{alg3}
 [\hat x_s,\hat y_s]&=&0,\,\,\,[\hat x_s,\hat p_{x_s}]=i\hbar,\,\,\,[\hat y_s,\hat  p_{y_s}]=i\hbar, \cr
[\hat p_{x_s},\hat  p_{y_s}]&=&0,\,\,\, [\hat x_s,\hat  p_{y_s}]=0,\,\,\,\,\,\,\, [\hat y_s, \hat p_{x_s}]=0.
\end{eqnarray}
In terms of the standard flat-Hermitian noncommutative  operators (\ref{alg2}),  we may now represent the algebra (\ref{alg1}) as follows
\begin{eqnarray}\label{T}
  \hat X=(1-\tau \hat y_0+\tau^2\hat y_0^2)\hat x_0,
  \quad  \hat Y=\hat y_0,\quad 
   \hat P_x=\hat p_{x_0},\quad 
   \hat P_y=(1-\tau \hat y_0+\tau^2\hat y_0^2)\hat p_{y_0}.
 \end{eqnarray}
 From this representation follows immediately that some of the operators involved are no
longer Hermitian. We observe
\begin{eqnarray}\label{key}
\hat X^\dag=\hat X-i\theta \tau (1-2\tau  \hat Y),\quad  \hat Y^\dag =\hat Y, \quad  \hat P_x^\dag= \hat P_x,\quad
 \hat P_y^\dag= \hat P_y+i\hbar\tau(\mathbb{I}-2\tau \hat Y).
\end{eqnarray}
As is apparent, the operators $\hat X$ and  $\hat P_y$ are not Hermitian. As an immediate consequence,  the Hamiltonian of the system involving these operators will in
general also not be Hermitian i.e. $\hat H^\dag(\hat X,\hat Y,\hat P_x,\hat P_y)\neq \hat H(\hat X,\hat Y,\hat P_x,\hat P_y)$.  In order  to map this operator into  Hermitian ones, some synonymously used concepts are introduced in the literature such as the $\mathcal{PT}$-symmetry \cite{29,30,31,32}, the  quasi-Hermiticity \cite{33,34}, the  pseudo-Hermiticity \cite{35,36,37} or the cryptoHermiticity \cite{38,39,40}. 
It has been clarified in \cite{30}  that a non-Hermitian operator  $\mathcal{O}$ having all eigenvalues real is connected to its Hermitian conjugate $\mathcal{O}^\dag$ through a linear, Hermitian, invertible, and bounded metric operator $\rho$ such  as $\rho \mathcal{O} \rho^{-1}= \mathcal{O}^\dag$. Factorizing this operator into
a product of a Dyson operator $\eta$ and its Hermitian conjugate in the form 
$\rho=\eta^\dag \eta$, it is  established \cite{30} that the non Hermitian operator   can be transformed to an equivalent Hermitian one given by $o=\eta \mathcal{O}\eta^{-1}=o^\dag$.
Schematically summarized, the latter can be described
by the following sequence of steps
\begin{eqnarray}
\mathcal{O}&\neq&  \mathcal{O}^\dag \xrightarrow{\rho} \rho \mathcal{O} \rho^{-1}= \mathcal{O}^\dag \xrightarrow{\eta} \eta \mathcal{O}\eta^{-1}=o=o^\dag.
\end{eqnarray}
For the case at hand, we find that the Dyson map can be taken to be
\begin{eqnarray}
 \eta=(1-\tau\hat Y +\tau^2 \hat Y^2)^{-1/2},
\end{eqnarray}
 so  the hidden Hermitian variables $\hat x, \hat y, \hat p_x , \hat p_y$ can be stated in terms of $\theta$-deformed
space operators as follows 
\begin{eqnarray} \label{non}
\hat x&=&\eta \hat X\eta^{-1}= (1-\tau\hat y_0 +\tau^2 \hat y_0^2)^{1/2}\hat x_0 (1-\tau\hat y_0 +\tau^2 \hat y_0^2)^{1/2}=\hat x^\dag,\label{a1}\\
 \hat p_x&=&\eta  \hat P_x\eta^{-1}=\hat p_{x_0}=\hat p_x^\dag,\label{p_1} \label{a2}\\
\hat y &=& \eta \hat Y\eta^{-1}=\hat y_0=y^\dag,\label{a3}\\
 \hat  p_y&=& \eta \hat P_y\eta^{-1}=  
 (1-\tau\hat y_0 +\tau^2 \hat y_0^2)^{1/2}\hat p_{y_0} (1-\tau\hat y_0 +\tau^2 \hat y_0^2)^{1/2}=\hat  p_y^\dag \label{p_2}.\label{a4}
\end{eqnarray} 
These operators satisfy the same deformed canonical
commutation relations as their counterparts in the non-Hermitian version of the theory (\ref{alg1})
\begin{eqnarray} \label{alg3}
[\hat x,\hat y]&=&i\theta (1-\tau\hat y +\tau^2 \hat y^2),\quad [\hat x,\hat p_x ]=i\hbar (1-\tau \hat y +\tau^2 \hat y^2),\cr
 {[\hat y,\hat p_y ]}&=&i\hbar (1-\tau \hat y +\tau^2 \hat y^2),\quad
[\hat x,\hat p_y]=i\hbar\tau(2\tau \hat y\hat x-\hat x)+
  i\theta\tau(2\tau \hat y\hat p_y-\hat p_y),\cr
 { [\hat p_x,\hat p_y]}&=&0, \quad \quad\quad \quad\quad \quad \quad\quad   {[\hat y, \hat p_x]}=0. 
\end{eqnarray}

As is well established in \cite{30}, a consequence of the non-Hermiticity of an operator 
$\mathcal{O}$, its eigenstates no longer form an orthonormal basis and the
Hilbert space representation has to be modified.
This is achieved by utilizing the operator $\rho$ as a metric to define a
new inner product $\langle . \,|\, .\rangle_\rho$ in terms of the standard inner
product  $\langle .\, |\,.\rangle$ defined as
\begin{eqnarray}\label{str1}
\langle \Phi | \Psi\rangle_\rho:=\langle \Phi|\rho \Psi\rangle,
\end{eqnarray}
for arbitrary states $\langle \Phi |$ and $ | \Psi\rangle$. The observables $\mathcal{O}$
are then Hermitian with respect to this new metric 
\begin{eqnarray}
\langle \Phi |\mathcal{O} \Psi\rangle_\rho=\langle \mathcal{O}\Phi|\rho \Psi\rangle.
\end{eqnarray}
An important physical consequence resulting from the   algebra (\ref{alg1}), is the loss of the Hermicity of certain operators which deformed the  structure of the Hilbert  space (\ref{str1}) 
as were predicted by the theory of Kempf {\it et al} \cite{7}. In the next section, let us study  Heisenberg’s uncertainty principle applied to a simultaneous measurement of operators of this algebra.

\section{Minimal and maximal measurements}\label{sec3}
\begin{figure}[htbp]
    	\resizebox{0.7\textwidth}{!}{
    		\includegraphics{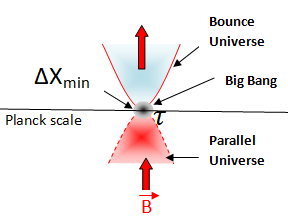}
    	 	  	 	  	 	} 	  	 	
    	\caption{\it \small Representation of minimal length scale $ \Delta X_{min}$.
    	}   	
    	\label{fig0}       
    \end{figure}

\begin{figure}[htbp]
      	\resizebox{0.9\textwidth}{!}{
      		\includegraphics{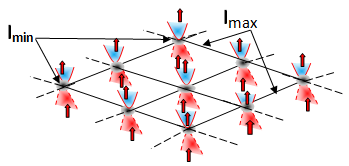}
      	 	  	 	  	 	} 	  	 	
      	\caption{\it \small A simultaneous representation of minimal length  $ \Delta X_{min}$ and maximal length  $ \Delta X_{max}$.
      	}   	
      	\label{fig01}       
      \end{figure}

For the system of operators satisfying the commutation relations in (\ref{alg1}), the generalized uncertainty principle is defined as follows
\begin{eqnarray}
\Delta  A\Delta  B\geq \frac{1}{2}|\langle [\hat A,\hat B]\rangle_\rho|\quad \mbox{for}
\quad  \hat A,\hat B\in \{\hat X,\hat Y, \hat P_x,\hat P_y\},
\end{eqnarray}
where $ \Delta  A=\sqrt{\langle (\hat A-\langle \hat A\rangle_\rho)^2\rangle_\rho} $ and for $\hat B$.
 An interesting features  can be observed through the  following  uncertainty relations:
\begin{eqnarray}
 \Delta  X\Delta Y&\geq&\frac{\theta}{2}\left(1-\tau\langle \hat Y\rangle_\rho +\tau^2\langle \hat Y^2\rangle_\rho\right),\label{in1}\\
 \Delta  Y\Delta P_y&\geq& \frac{\hbar}{2}\left(1-\tau\langle \hat Y\rangle_\rho+\tau^2\langle \hat Y^2\rangle_\rho\right)\label{in2}.
 \end{eqnarray}
 $i)$  For the uncertainty relation (\ref{in1}), using 
 $ \langle \hat Y^2\rangle_\rho =\Delta  Y^2+\langle \hat Y\rangle_\rho^2$, the inequality (\ref{in1}) is reduced to 
 \begin{eqnarray}
 \Delta  X\Delta Y&\geq&\frac{\theta}{2}\left(1 +\tau^2 \Delta Y^2\right)\label{A}\quad \mbox{for}\quad  \langle \hat Y\rangle_\rho=0.
 \end{eqnarray}
 This equation (\ref{A}) can be rewritten as a second order equation of $ \Delta Y $  and the  solutions are given by

\begin{equation}
\Delta X=\frac{\Delta Y}{\theta \tau^2}\pm \sqrt{\left(\frac{\Delta Y}{\theta \tau^2}\right)^2
	-\frac{1}{\tau^2}}.
\end{equation}
These solutions lead to the absolute minimal uncertainty $ \Delta X_{min} $ in $X$-direction and to the absolute maximal uncertainty $ \Delta Y_{max} $ in $Y$-direction as  predicted by Perivolaropoulos \cite{1}
\begin{eqnarray}
\Delta X_{min}&=&\theta\tau=\frac{\tau}{B}=l_{min},\\
\Delta Y_{max}&=&\frac{1}{\tau}=l_{max}.\label{max}
\end{eqnarray}
Different versions of minimal length uncertainties  have been  introduced in the literature \cite{50,51,52,53} which  significantly improve the  one proposed by  Kempf et al  \cite{7}. It is  well known that these  minimal length uncertainties induce a singularity of position representation at the Planck scale  i.e they are inevitably bounded by minimal
quantities beyond which any further localization of particles is not possible. Conversely  to these results, here the obtained minimal length  $ \Delta X_{min}=\frac{\tau}{B}$ induces a broken singularity at the Planck  scale due to the external   magnetic fields $B$. In fact this scenario can be regarded as the Landau problem where the Planck scale bounded by the weak quantum gravitational field $\tau$ is orthogonally subjected to the superstrong magnetic field $B$ of the parallel universe causing  its bouncing at this  minimal point. This broken singulary manifested by a big bang unifies the weak quantum  gravitational  field and the superstrong magnetic field as minimal length (see Figure \ref{fig0}).
A simultaneous measurement of  the minimal length $ \Delta X_{min}$  and the maximal length $ \Delta Y_{max}$ generates the inverse of  the magnetic field as follows\label{inversemagnetic}
\begin{eqnarray}\label{eqq22}
\Delta X_{min} \Delta Y_{max}=\frac{1}{B}=l_{min}l_{max}.
\end{eqnarray}
If we consider $n$-dimensional sets of the algebra (\ref{alg1}), based on the equation (\ref{eqq22}) we obtain a sequence of minimal lengths alternated by maximal lengths
\begin{eqnarray}
...l_{min}l_{max}l_{min}l_{max}...=\frac{1}{B^n}.
\end{eqnarray}
This  sequence forms a sort of discreteness
of the Planck space and  can be compared to a lattice system in which each site represented by $l_{min}$ is spaced by $l_{max}$ (see Figure \ref{fig01}).
  Note that two  of the strongest competing candidates for a theory of quantum gravity, String Theory (ST) and Loop Quantum Gravity (LQG) have been thought of as
  describing different regimes of  quantum gravity.   ST fundamentally attempts to unify  forces while LQG attempts to quantize  spacetime at the Planck scale. Based on the proposal (\ref{eqq22})  and  (Figure \ref{fig01}), we can argue that ST and LQG are in fact not  fundamentally different from each other, they are just two different methods to approach   the same problem.  \\
ii) Repeating the same calculation  and argumentation in the situation of  uncertainty relation (\ref{max}) for simultaneous $\hat Y,\hat P_y$-measurement, we find
the absolute  maximal uncertainty  $\Delta Y_{max}$ (\ref{max}) and an absolute minimal uncertainty momentum $\Delta P_{y_{min}}$  for  $\langle  \hat Y\rangle_\rho=0$
\begin{equation}
 \Delta Y_{max}=\frac{1}{\tau}=l_{max},\quad \Delta P_{y_{min}}=\hbar \tau=p_{min}.\label{pmin}
\end{equation}
These results (\ref{pmin})
are consistent with the ones obtained by  Perivolaropoulos \cite{1}. From these results,
it is interesting to observe that the GUP is reduced into 
 $\Delta Y_{max}\Delta P_{y_{min}}=\hbar$. It is
well known from the Heisenberg principle that the latter relation can be cast into
\begin{equation}\label{eq25}
\Delta Y_{max}\Delta E=\hbar c\implies\Delta E=\frac{\hbar c }{\Delta Y_{max}}.
\end{equation}
where $\Delta E= \Delta P_{y_{min}} c$. Unlike the results obtained in the minimal length scenarios
\cite{7,10,49',49'',49''',49'''',50,51,52}, here the required uncertainty energy is weak since the dimension
of length $ \Delta Y_{max} $ is very large. This indicates that a maximal localization of quantum gravity induces weak energies for its measurement. Let us now consider
the equation $ \Delta Y_{max} =\Delta t c$. Inserting this equation in (\ref{eq25}), one obtains
\begin{equation}
\Delta t=\frac{\hbar }{\Delta E}.
\end{equation}
Since, the uncertainty energy is low in this space due to the maximal measurement of quantum gravity, then its time $ \Delta t$ strongly increases i.e the time runs more
slowly in this space. In comparison with the minimal length theories, the concept
of a maximal length of quantum gravity developed in this paper admits a close analogy with
the properties of gravity in GR in the sense that the gravitational field becomes stronger for heavy systems that curve the space,
allowing  the  surrounding  light systems to fall down with low energies.  The resulting time inside of this space is dilated and length  contraction takes effect. As will be shown in the next section, the increase of the quantum
gravitational parameter $\tau$ in an infinite square well potential curves the quantum
levels to enable enable the particles to jump from one state to another with low energy  and with high probability densities. The wavefunction  compresses and contracts inward as
one increases the effect of quantum gravitational effects. \\
iii)  Finally, simultaneous measurements of operators $(\hat X, \hat P_y)$, $(\hat X, \hat P_x)$ and $ (\hat X, \hat P_y)$
are spatial isotropy since their measurements do not present any minimal/maximal length or minimal momentum.

\begin{center}
   \begin{figure}[htbp]
    	\resizebox{0.9\textwidth}{!}{
    		\includegraphics{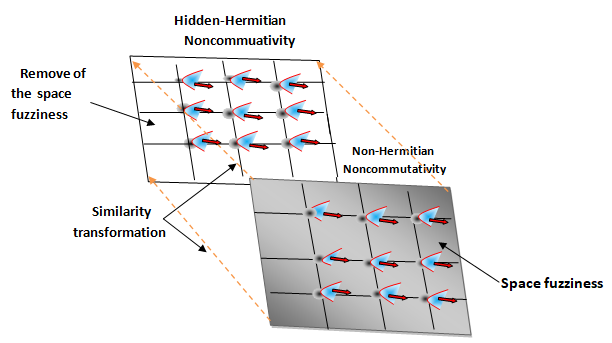}
    	 	  	 	  	 	}	 	  	 	
    	\caption{\it \small Hidden Hermitian noncommutativity and Non-Hermitian noncommutativity.
    	}
    	
    	\label{fig1}       
    \end{figure}
  
\end{center}

Moreover, repeating the GUP calculations with the hidden position-dependent noncommutativity (\ref{alg3}), one generates the same uncertainty measurements 
\begin{eqnarray}
\Delta x_{min}=\theta\tau,\quad \Delta x_{max}=\frac{1}{\tau},\quad \Delta p_{min}=\hbar \tau.
\end{eqnarray}
This indicates that the Dyson map does not remove the characteristics of  quantum gravity at this   scale i.e it only removes the fuzziness induced by the singular points by shedding light on the hidden Hermitian space. Consequently particles can be localized in precise ways in this new space. This  situation could be compared to the  gravitational holographic principle where the real information inside the Black hole is virtually projected at its event horizon. Taking into account the latter, the formal   simultaneous representation of minimal length  $ \Delta X_{min}$ and maximal length  $ \Delta X_{max}$ (Figure  \ref{fig01}) can be illustrated as follows (Figure \ref{fig1})

\section{ Hidden-Hermitian free particle Hamiltonian
	in a box} \label{sec4}
\begin{center}
   \begin{figure}[htbp]
    	\resizebox{0.9\textwidth}{!}{
    		\includegraphics{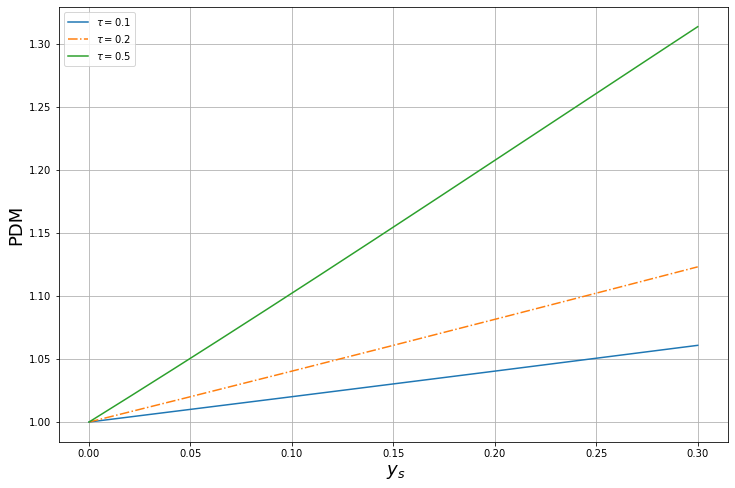}
    	 	  	 	  	 	}	 	  	 	
    	\caption{\it \small PDM  versus the position $y_s$ for different values of $\tau$.
    	}
    	
    	\label{figPDM}       
    \end{figure}
  
\end{center}

The Hamiltonian of free particle in 2D non-Hermitian position-dependent noncommutative space reads as follows
\begin{eqnarray} \label{Hami}
\hat H_F=\frac{1}{2m_0}\left(\hat P_x^2+\hat P_y^2\right).
\end{eqnarray}
As mentioned, any Hamiltonian depending on the operators $\hat X$ or $\hat P_y$ will obviously
no longer be Hermitian. Thus, using the relations (\ref{T}), we can transform the Hamiltonian (\ref{Hami}) into the standard  $\theta$-deformed operator (\ref{alg2}) as follows
\begin{eqnarray}
\hat H_F=\frac{1}{2m_0}\left[\hat p_{x_0}^2+\left(1-\tau \hat y_0+\tau^2 \hat y_0^2\right)^2 \hat p_{y_0}^2-i\hbar \tau (1+2\tau \hat y_0)(1-\tau \hat y_0 +\tau^2\hat y_0^2)\hat p_{y_0}\right].
\end{eqnarray}
Evidently this Hamiltonian is non-Hermitian $\hat H\neq\hat H^\dag$. Thus, the Hermicity requirement  of this operator  is achieved by means of  a similarity transformation. Since all our variables are converted into Hermitian ones by the same Dyson map, this will also hold for any function in these variables, as for instance the Hamiltonian. Thus, the Hermitian counterpart Hamiltonian becomes
\begin{equation}
\hat h_F=\eta \hat H_F\eta^{-1}= \frac{1}{2m_0}\left(\hat p_x^2+\hat p_y^2\right).
\end{equation}
Using  the relation (\ref{p_1},\ref{p_2}), we rewrite the Hamiltonian in terms of the $\theta$-noncommutative operators
\begin{eqnarray}
\hat h_F=\frac{1}{2m_0}\left[\hat p_{x_0}^2+  (1-\tau\hat y_0 +\tau^2 \hat y_0^2)^{1/2}\hat p_{y_0} (1-\tau\hat y_0 +\tau^2 \hat y_0^2) \hat p_{y_0} (1-\tau\hat y_0 +\tau^2 \hat y_0^2)^{1/2}\right].
\end{eqnarray}
Appealing to the nonsymmetric Bopp-shift (\ref{bob}), we may  rewrite, the above Hamiltonian as follows
\begin{eqnarray}
\hat h_F=\frac{1}{2m_0}\left[\hat p_{x_s}^2+  (1-\tau\hat y_s +\tau^2 \hat y_s^2)^{1/2}\hat p_{y_s} (1-\tau\hat y_s +\tau^2 \hat y_s^2) \hat p_{y_s} (1-\tau\hat y_s +\tau^2 \hat y_s^2)^{1/2}\right].
\end{eqnarray}
The  time-independent  Schrödinger equation is given by
\begin{eqnarray}\label{schro}
\hat h_F\psi(x_s,y_s)&=& E\psi(x_s,y_s),\cr
(\hat h_F^x+\hat h_F^y) \psi(x_s,y_s)&=& E\psi(x_s,y_s).\label{schro}
\end{eqnarray}
As it is clearly seen, the system is decoupled and the solution to the eigenvalue equation (\ref{schro})
is given by
\begin{eqnarray}
\psi(x_s,y_s)= \psi(x_s)\psi(y_s), \quad E=E_x+E_y
\end{eqnarray}
where $\psi(x_s) $ is the wave function in the $x_s$-direction and $\psi(y_s) $  the wave function in the
$y_s$-direction. Since the particle is free in the $x_s$-direction, the wave function is given by \cite{25}
\begin{eqnarray}
\psi_k(x_s)=\int_{-\infty}^{+\infty} dk g(k)e^{ikx_s},
\end{eqnarray}
where $g(k)$ determines the shape of the wave packet and the energy spectrum is continuous \cite{25}
\begin{eqnarray}
E_x=E_k=\frac{\hbar^2 k^2}{2m_0}.
\end{eqnarray}
In $y_s$-direction, we have to solve  the following equation 
\begin{eqnarray}
\frac{1}{2m_0} (1-\tau\hat y_s +\tau^2 \hat y_s^2)^{1/2}\hat p_{y_s} (1-\tau\hat y_s +\tau^2 \hat y_s^2) \hat p_{y_s} (1-\tau\hat y_s +\tau^2 \hat y_s^2)^{1/2}\psi(y_s)=E_y\psi(y_s).
\end{eqnarray}
This equation  is an agreement with the one introduced by von Roos \cite{55} for systems with a  position-dependent mass (PDM) operator  and it can be rewritten as \cite{55'}
\begin{eqnarray}\label{eq22}
\left(-\frac{\hbar^2}{2m_0}\sqrt[4\,]{\frac{m_0}{m(y_s)}}\frac{\partial}{\partial_{y_s}}\sqrt{\frac{m_0}{m(y_s)}}\frac{\partial}{\partial_{y_s}}\sqrt[4\,]{\frac{m_0}{m(y_s)}}\right)\psi(y_s)=E_y \psi(y_s),	
\end{eqnarray}
where
\begin{eqnarray}
m(\hat  y_s)=\frac{m_0}{(1-\tau\hat y_s +\tau^2 \hat y_s^2)^2 },
\end{eqnarray}
being   the PDM of the  system strongly pertubated by quantum gravity \cite{12'}. Figure \ref{figPDM}  illustrates
the PDM as a function of the position $y_s$ $(0 < y_s < 0,3)$. One observes that the effective
mass $m(\hat  y_s)$ in this description increases with $\tau$. This indicates that quantum gravitational
fields increase with  $ m(\hat  y_s)$. 
In otherwise, by increasing experimentally the PDM, one can make the quantum gravitational effects stronger for a measurement  through the variation of PDM of the system. Furthermore, the increase of PDM with the quantum gravitational effect  will be consequence of the  deformation  of the quantum energy levels  allowing, the particle to jump from one state to another with low energies (\ref{fig5}).
 This observation  is in
perfect analogy with the theory of GR where massive objects induce strong gravitational fields and curve the space enabling the  the surronding light systems to fall down with low energies.

The equation (\ref{eq22}) can be conveniently rewritten by means of the transformation $\psi(y_s)=\sqrt[4\,]{m(y_s)/m_0}\phi(y_s)$ as in \cite{55'}
\begin{eqnarray}
-\frac{\hbar^2}{2m_0}\left(\sqrt{\frac{m_0}{m(y_s)}}\frac{\partial}{\partial_{y_s}}\right)^2\phi(y_s)=E_y \phi(y_s),\quad \mbox{with} \quad E_y >0,	
\end{eqnarray}
	or 
\begin{eqnarray}\label{eq1}
-\frac{\hbar^2}{2m_0}\left[(1-\tau  y_s+\tau^2 y_s^2)^2\frac{\partial^2}{\partial_{y_s^2}}- \tau(1-2\tau y_s)(1-\tau  y_s+\tau^2 y_s^2)\frac{\partial}{\partial_{y_s}}\right] \phi (y_s)=E_y \phi (y_s).
\end{eqnarray}
The solution of this equation (\ref{eq1}) is  given by \cite{12'}
\begin{eqnarray}\label{wav}
\phi_\lambda(y_s)&=& A\exp\left(i\frac{2\lambda}{\tau\sqrt{3}}
\left[\arctan\left(\frac{2\tau y_s-1}{\sqrt{3}}\right)+\frac{\pi}{6}\right]\right),\\
\psi_\lambda(y_s)&=&\frac{A}{\sqrt{1-\tau y_s+\tau^2 y_s^2}}\exp\left(i\frac{2\lambda}{\tau\sqrt{3}}
\left[\arctan\left(\frac{2\tau y_s-1}{\sqrt{3}}\right)+\frac{\pi}{6}\right]\right),
\end{eqnarray}
where $\lambda=\frac{\sqrt{2m_0E_y}}{\hbar} $ and A is the normalization constant. We notice that if the standard wave-function $\psi_\lambda(y_s)$ is normalized,
then $\phi_\lambda(y_s)$ is normalized under a $\tau$-deformed integral. Indeed, we have
\begin{eqnarray}\label{eq4}
\int_{-\infty}^{+\infty} dy_s\psi_\lambda^*(y_s)\psi_\lambda(y_s)= \int_{-\infty}^{+\infty}\frac{dy_s}{1-\tau y_s+\tau^2 y_s^2} \phi_\lambda^*(y_s)\phi_\lambda(y_s)=1.
\end{eqnarray}
 Based on this   equation (\ref{eq4}), the  normalized  constant $A$ is determined  as follows 
  \begin{eqnarray}
  1&=&\int_{-\infty}^{+\infty}\frac{dy_s}{1-\tau y_s+\tau^2 y_s^2} \phi_\lambda^*(y_s)\phi_\lambda(y_s)\\
  &=& A^2 \int_{-\infty}^{+\infty}\frac{dy_s}{1-\tau y_s+\tau^2 y_s^2},
  \end{eqnarray}
so, we find
\begin{eqnarray}\label{nzeta}
A=\sqrt{\frac{\tau\sqrt{3}}{2\pi}}. 
\end{eqnarray}
The next important point concerns, is the quantization of the energy spectrum; we will show below that this property comes directly from the orthogonality of these solutions.  Since the operator $\hat h_y$ is  Hermitian, then the corresponding   eigenfunctions $\psi_\lambda(y_s)$ are  orthogonal. This property can be shown by considering the integral 
\begin{eqnarray}\label{nor}
\int_{-\infty}^{+\infty} dy_s\hat h_y^\dag\psi_{\lambda'}^*(y_s)\psi_{\lambda}(y_s)= \int_{-\infty}^{+\infty} dy_s\hat \psi_\lambda^*(y_s)\hat h_y\psi_\lambda(y_s),
\end{eqnarray}
which becomes, after an integration by parts,
\begin{eqnarray}
E_{\lambda'}\int_{-\infty}^{+\infty} dy_s \psi_{\lambda'}^*(y_s)\psi_{\lambda}(y_s)= E_{\lambda} \int_{-\infty}^{+\infty} dy_s\hat \psi_{\lambda'}^*(y_s)\psi_\lambda(y_s).
\end{eqnarray}
Since these two integrals are equal, one has 
\begin{eqnarray}
(E_{\lambda'}-E_{\lambda})\int_{-\infty}^{+\infty} dy_s \psi_{\lambda'}^*(y_s)\psi_{\lambda}(y_s)&=&0,\\
(E_{\lambda'}-E_{\lambda})\int_{-\infty}^{+\infty} dy_s \frac{A^2}{1-\tau y_s+\tau^2 y_s^2}e^{i\frac{2(\lambda-\lambda')}{\tau\sqrt{3}}
\left[\arctan\left(\frac{2\tau y_s-1}{\sqrt{3}}\right)+\frac{\pi}{6}\right]}&=&0,\\
\sin\left(\frac{\lambda-\lambda'}{\tau\sqrt{3}}\pi\right)&=&0. \label{eq}
\end{eqnarray}
The quantization follows from the equation (\ref{eq}) and leads to the equation
\begin{eqnarray}
\frac{\lambda-\lambda'}{\tau\sqrt{3}}\pi&=&n\pi \cr
\lambda-\lambda'&=& \lambda_n=\tau n\sqrt{3}, \quad \quad \quad n\in \mathbb{N},
\end{eqnarray}
where one notices the case $n=0$  i.e. $\lambda=\lambda' $, corresponding to the normalization condition considered in (\ref{nor}). Then, the energy spectrum of the particle is written as
\begin{eqnarray}
E_y=E_n=\frac{3\tau^2 \hbar^2}{2m_0}n^2.
\end{eqnarray}
As it is clairly obtained, the presence of this  deformed parameter $\tau$ in $y_s$-direction quantized the energy of a free particle. This fact comes to confirm the fundamental property of gravity which consists of contracting  and discretizing the matter.
 
Then,  the total eigensystem is given by 
\begin{eqnarray}
E=\begin{cases}
E_x= \frac{\hbar^2 k^2}{2m_0},\\
E_y=\frac{3\tau^2 \hbar^2}{2m_0}n^2.
\end{cases}
\end{eqnarray}
and 
\begin{eqnarray}
\psi(x_s,y_s)=\begin{cases}
\psi_k(x_s)= \int_{-\infty}^{+\infty} dk g(k)e^{ikx_s}, \\
\psi_n(y_s)= \frac{A}{\sqrt{1-\tau y_s+\tau^2 y_s^2}}\exp\left(i 2n
\left[\arctan\left(\frac{2\tau y_s-1}{\sqrt{3}}\right)+\frac{\pi}{6}\right]\right).
\end{cases}
\end{eqnarray}

\begin{figure}[htbp]
 	\resizebox{1\textwidth}{!}{
 		\includegraphics{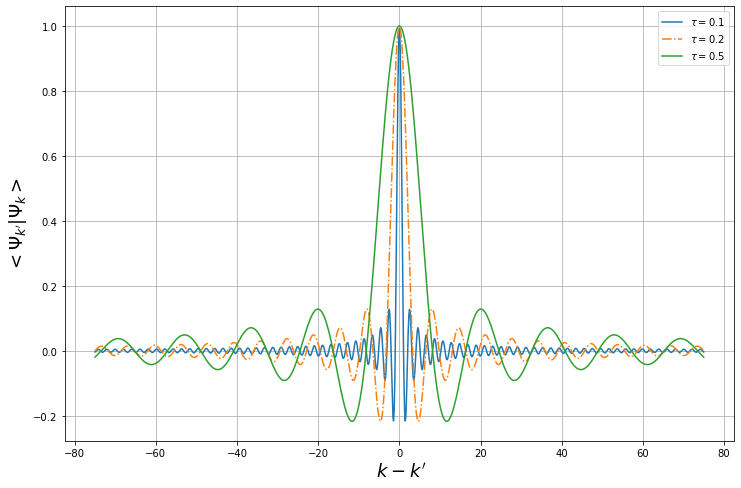}
 	}
 	\caption{\it \small Variation of $ \langle \psi_{k'}|\psi_k\rangle$ versus $k-k'$ with $a=1$. 
 	}
 	
 	\label{fig2}       
 \end{figure}

\begin{figure}[htbp]
 	\resizebox{1\textwidth}{!}{
 		\includegraphics{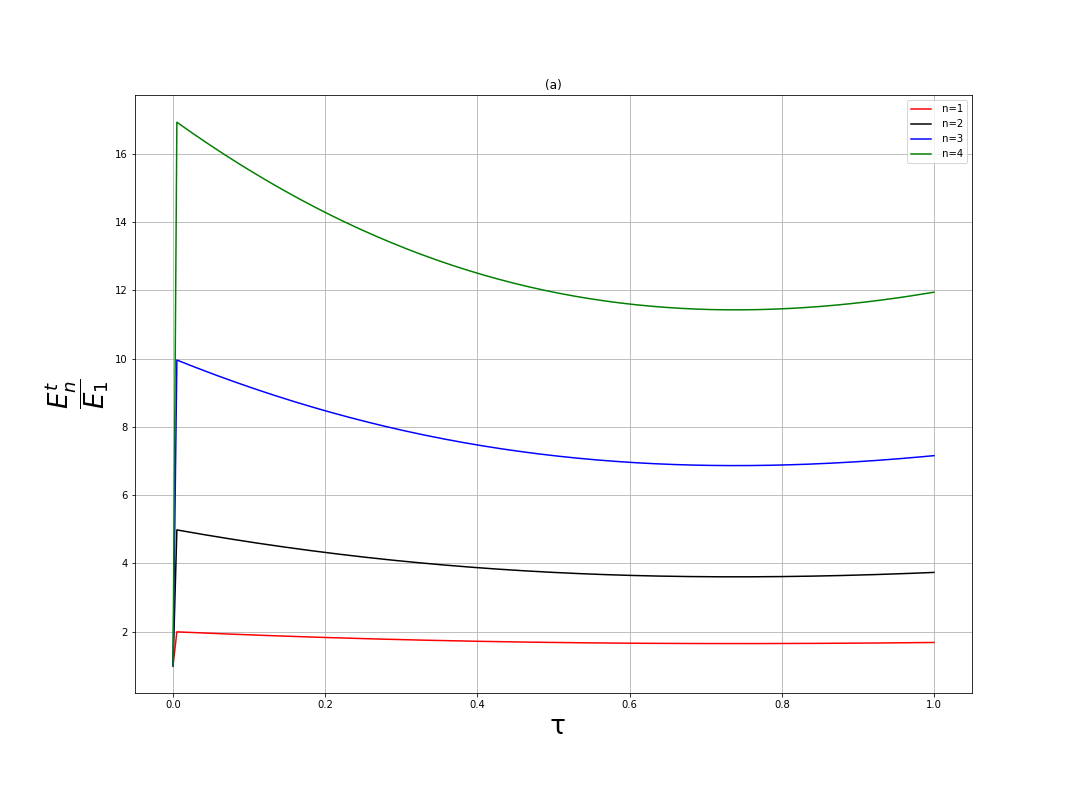}
 	}
 	\resizebox{1\textwidth}{!}{
 	 		\includegraphics{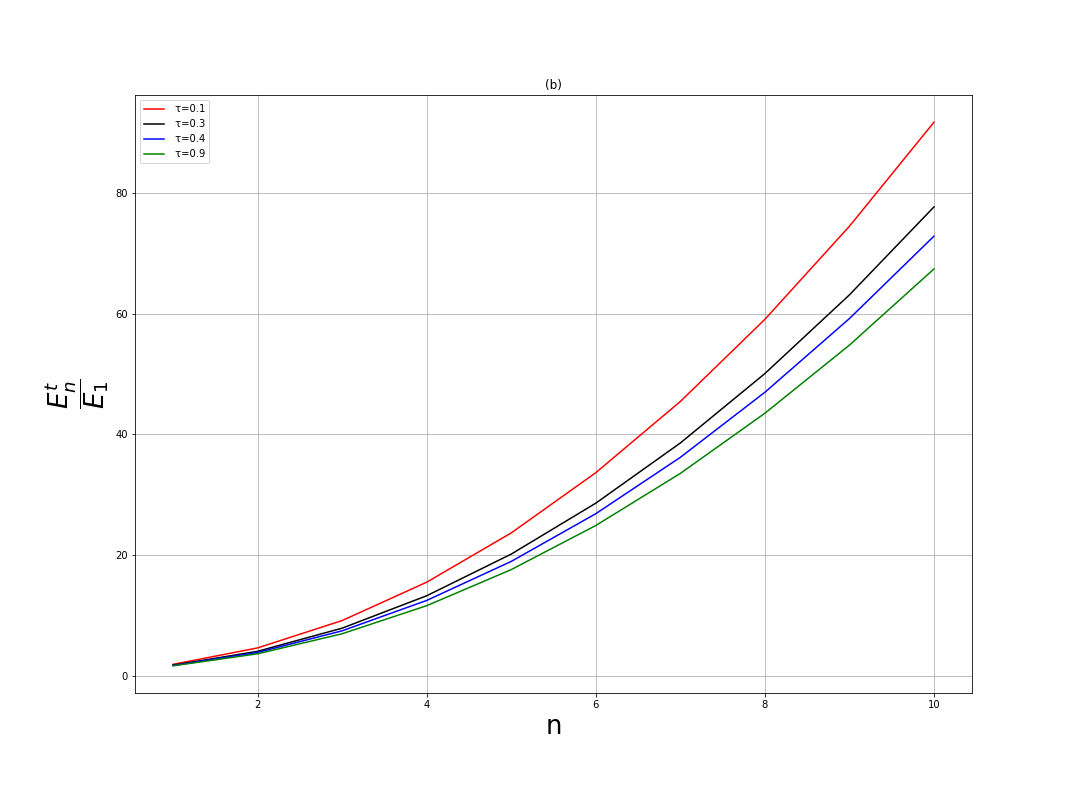}
 	 	}
 	\caption{\it \small   The energy $E_n/E_1$ of the particle in 2D box of length $a = 1$ with mass $m_0 = 1$ and
 	$\hbar = 1$.
 	}
 	
 	\label{fig5}       
 \end{figure}

 Now, we consider the above free particle of mass $m_0$ captured in a two-dimensional box of length $0\le x_s\le a$
and heigth $0\le y_s\le a$. The boundaries of the box are located. We impose the wave functions $\psi(0)=0=\psi(a)$.
The eigensystems in  $x_s$-direction are given by 
\begin{eqnarray}\label{eq8}
\psi_n(x_s)=\sqrt{\frac{2}{a}}\sin\left(\frac{n\pi}{a}x_s\right), \quad E_n=n^2\frac{\pi^2\hbar^2}{2m_0a^2},\quad E_1=\frac{\pi^2\hbar^2}{2m_0a^2}.
\end{eqnarray}
Taking the results (\ref{eq8}) as a witness, we study what follows the influence of the deformed
parameter $\tau$ on the system. In  $y_s$-direction, the solution is given by 
\begin{eqnarray}\label{wav}
\psi_k(y_s)=\frac{B}{\sqrt{1-\tau y_s+\tau^2 y_s^2}}\exp\left(i\frac{2k}{\tau\sqrt{3}}
\left[\arctan\left(\frac{2\tau y_s-1}{\sqrt{3}}\right)+\frac{\pi}{6}\right]\right),
\end{eqnarray}
where $k=\frac{\sqrt{2m_0E'}}{\hbar}$.  Then by normalization, $\langle \psi_k|\psi_k\rangle=1$,  we have 
\begin{eqnarray}
1&=& B^2\int_{0}^{a}\frac{dy_s}{1-\tau y_s+\tau^2 y_s^2},
\end{eqnarray}
so we find
\begin{eqnarray}\label{nzeta}
B=\sqrt{\frac{\tau\sqrt{3}}{2}} \left[\arctan\left(\frac{2\tau a-1}{\sqrt{3}}\right)+\frac{\pi}{6}\right]^{-1/2}.
\end{eqnarray}
Based on the reference \cite{7}, the scalar product of the formal eigenstates is given by
\begin{eqnarray}
\langle \psi_{k'}|\psi_k\rangle&=&\frac{\tau\sqrt{3}}{2(k-k')\left[\arctan\left(\frac{2\tau a-1}{\sqrt{3}}\right)\right]}\sin\left(\frac{2(k-k')\left[\arctan\left(\frac{2\tau a-1}{\sqrt{3}}\right)\right]}{\tau\sqrt{3}}\right).
\end{eqnarray}
This relation  shows that, the normalized  eigenstates (\ref{wav}) are  no longer
orthogonal. However,  if one tends $(k-k')\rightarrow \infty$, these states  become orthogonal 
\begin{eqnarray}
\lim_{(k-k')\rightarrow \infty} \langle \psi_{k'}|\psi_{k}\rangle=0.
\end{eqnarray}
These properties show that, the states
$|\psi_k\rangle$ are essentially Gaussians centered at $(k-k')\rightarrow 0$ (see Figure \ref{fig2}). This observation indicates   primodial fluctuations at this scale and these fluctuations increase with the quantum gravitational effects. As will be shown in the forthcoming development, the increase of fluctuations is manifested by  high probability densities of particles.
The states $|\psi_k\rangle$ can be compared
to the coherent states of harmonic oscillator \cite{55',55'',55'''} which are known as states 
that mediate a smooth transition between the quantum
and classical worlds. This transition is manifested by
the saturation of the Heisenberg uncertainty principle
$ \Delta_z q\Delta_z p=\hbar/2$. In comparison with coherent states of harmonic oscillator, the states $|\psi_{k}\rangle$  strongly saturate the GUP ($ \Delta_{\psi_{k}} X\Delta_{\psi_{k}}P=\hbar$) at
the Planck scale and could be used to describe the transition states between the quantum world and unknown
world for which the physical descriptions are out of reach.

We suppose  that, the wave function satisfies the  Dirichlet condition i.e it vanishes at the boundaries    $\psi_k(0)=0=\psi_k(a)$.
Thus, using especially the boundary condition $\psi_k(0)=0$,  the 
above wavefunctions (\ref{wav}) becomes
\begin{eqnarray}
\psi_k (y_s) =  \frac{B}{\sqrt{1-\tau y_s+\tau^2 y_s^2}}\sin\left(\frac{2k}{\tau \sqrt{3}}\left[\arctan\left(\frac{2\tau y_s-1}{\sqrt{3}}\right)
+\frac{\pi}{6}\right]\right).
\end{eqnarray}
The quantization follows from the boundary condition $\psi_k(a)=0$ and leads to the equation

\begin{eqnarray}
\frac{2k_n}{\tau \sqrt{3}}\left[\arctan\left(\frac{2\tau a-1}{\sqrt{3}}\right)
+\frac{\pi}{6}\right]&=&n\pi \quad \,\,\mbox{with}\quad n\in \mathbb{N}^*,\\
k_n &=&\frac{\pi\tau\sqrt{3}n}{2\left[\arctan\left(\frac{2\tau a-1}{\sqrt{3}}\right)
	+\frac{\pi}{6}\right]}.
\end{eqnarray}
Then, the  energy spectrum of the particle  is written as
\begin{eqnarray}\label{enrga}
E_n' &=& \frac{3\pi^2 \tau^2\hbar^2n^2}{8m_0 \left[\arctan\left(\frac{2\tau a-1}{\sqrt{3}}\right)
	+\frac{\pi}{6}\right]^{2} }.
\end{eqnarray}
At  the limit $\tau\rightarrow 0$, we have 
\begin{eqnarray}
\lim_{ \tau\rightarrow 0} E_n'= E_n=\frac{\pi^2\hbar^2n^2}{2m_0a^2}.
\end{eqnarray}
Thus, the energy levels can be rewritten as
\begin{eqnarray} \label{enrg2}
E_n=\frac{3}{4} \left[\frac{\tau L}{\arctan\left(\frac{2\tau L-1}{\sqrt{3}}\right)
	+\frac{\pi}{6} }\right]^2 E_n<E_n.
\end{eqnarray} 
The effects of the parameter $\tau$ in $y_s$ direction induce deformations of quantum levels, which consequently lead to a decrease in the amplitude of the energy levels.
The corresponding wave functions   to  the energies (\ref{enrga}) are given by
\begin{eqnarray}
\psi_n (x) = \frac{B}{\sqrt{1-\tau y_s+\tau^2 y_s^2}}\sin\left(\frac{ n\pi }{\left[\arctan\left(\frac{2\tau a-1}{\sqrt{3}}\right)
	+\frac{\pi}{6}\right]^{2}}\left[\arctan\left(\frac{2\tau y_s-1}{\sqrt{3}}\right)
+\frac{\pi}{6}\right]\right).\label{wa}
\end{eqnarray}
The total eigenvalues of the system are given by

\begin{eqnarray}\label{enrg1}
E_n^t &=& n^2\frac{\pi^2\hbar^2}{2m_0a^2}+ \frac{3\pi^2 \tau^2\hbar^2n^2}{8m_0 \left[\arctan\left(\frac{2\tau a-1}{\sqrt{3}}\right)
	+\frac{\pi}{6}\right]^{2} },\cr
&=&\left(1+\frac{3\tau^2 a}{\left[\arctan\left(\frac{2\tau a-1}{\sqrt{3}}\right)
	+\frac{\pi}{6}\right]^{2}}\right)E_n,
\end{eqnarray}
and 
\begin{eqnarray}\label{enrg1}
\lim_{ \tau\rightarrow 0}  E_n^t =2E_n,
\end{eqnarray}
The  wave function in $x_s,y_s$-directions are given by
\begin{eqnarray}
\psi (x_s,y_s) &=& \frac{B\sqrt{\frac{2}{a}}}{\sqrt{1-\tau y_s+\tau^2 y_s^2}} \sin\left(\frac{n\pi}{a}x_s\right)\cr&&\times\sin\left(\frac{ n\pi }{\left[\arctan\left(\frac{2\tau a-1}{\sqrt{3}}\right)
	+\frac{\pi}{6}\right]^{2}}\left[\arctan\left(\frac{2\tau y_s-1}{\sqrt{3}}\right)
+\frac{\pi}{6}\right]\right).\label{wa}
\end{eqnarray}
At  the limit $\tau\rightarrow 0$, we have 
\begin{eqnarray}
\lim_{ \tau\rightarrow 0} \psi (x_s,y_s)= \frac{2}{a} \sin\left(\frac{n\pi}{a}x_s\right)\sin\left(\frac{n\pi}{a}y_s\right).
\end{eqnarray}
The corresponding probability density is given by
\begin{eqnarray}
\rho (x_s,y_s) &=& \frac{2B}{a(1-\tau y_s+\tau^2 y_s^2)} \sin^2\left(\frac{n\pi}{a}x_s\right)\cr&&\times\sin\left(\frac{ n\pi }{\left[\arctan\left(\frac{2\tau a-1}{\sqrt{3}}\right)
	+\frac{\pi}{6}\right]^{2}}\left[\arctan\left(\frac{2\tau y_s-1}{\sqrt{3}}\right)
+\frac{\pi}{6}\right]\right).\label{wa}
\end{eqnarray}
Figure \ref{fig5} illustrates the energy levels of the particle as functions of the quantum number $n$ and the quantum gravitational parameter $\tau$. Figure (a) shows how the increase
of $\tau$ gradually curves the energy levels as one increases the quantum number $n$ from the fundamental level. Figure (b) shows  energy levels versus the quantum number $n$ for fixed values of $\tau$. Conversely to the graph obtained in \cite{55',57,58,59}, one observes that, the amplitudes of energy levels $E_n^t/E_1$ decrease
when $\tau$ increases. In fact, increasing the quantum gravitational effects  lead to the enhancement of binding
quantum levels  allowing particles to jump from one state to another with low energies \cite{12'}.

\begin{figure}[htbp]
  	\resizebox{0.5\textwidth}{!}{
  		\includegraphics{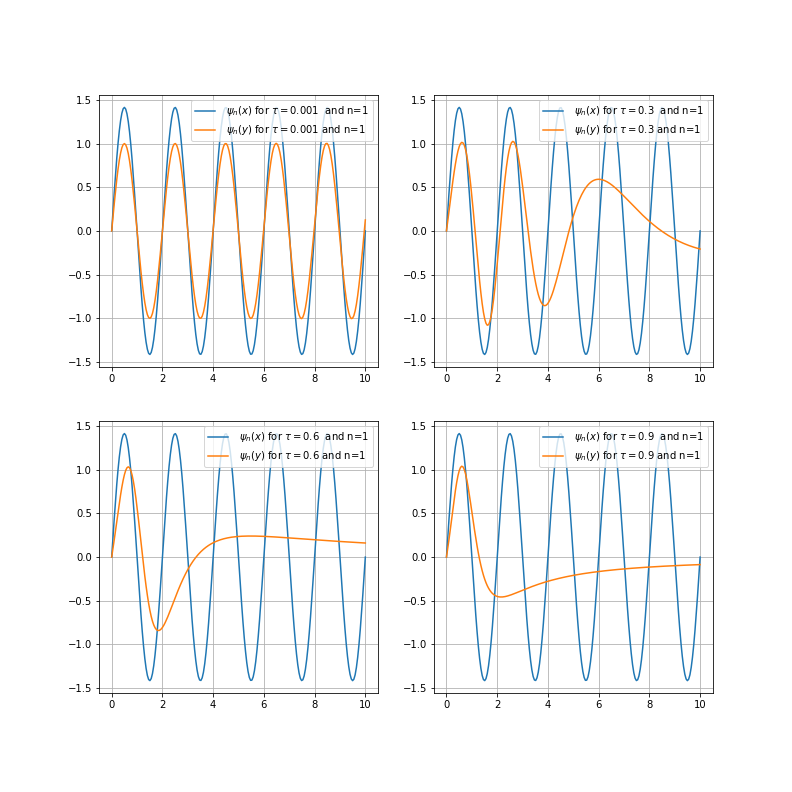}
  	}
  	\resizebox{0.5\textwidth}{!}{
  	 		\includegraphics{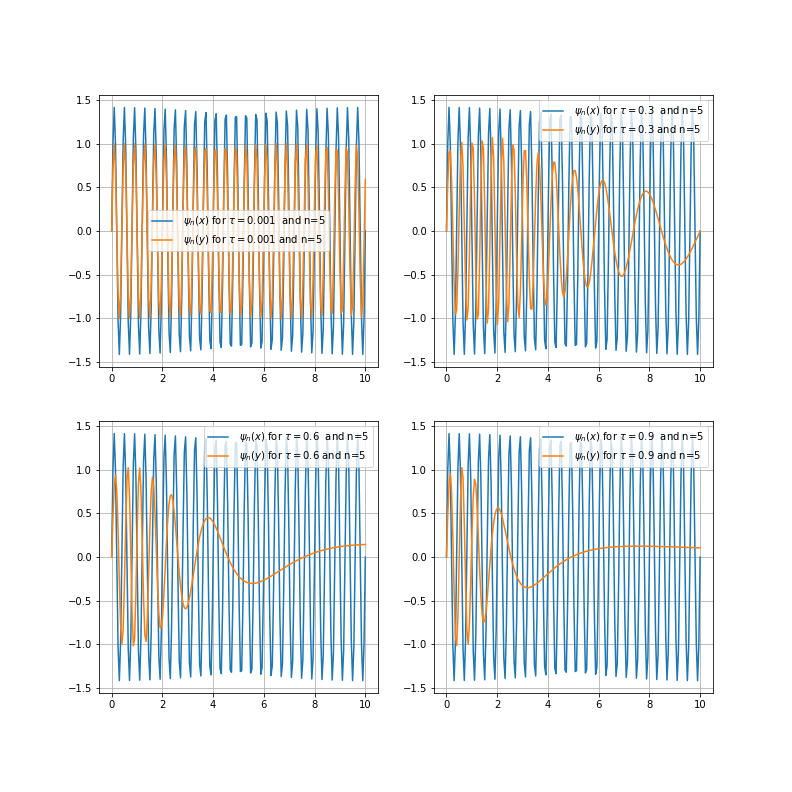}
  	 	}
  	 		\resizebox{0.5\textwidth}{!}{
  	 	  	 		\includegraphics{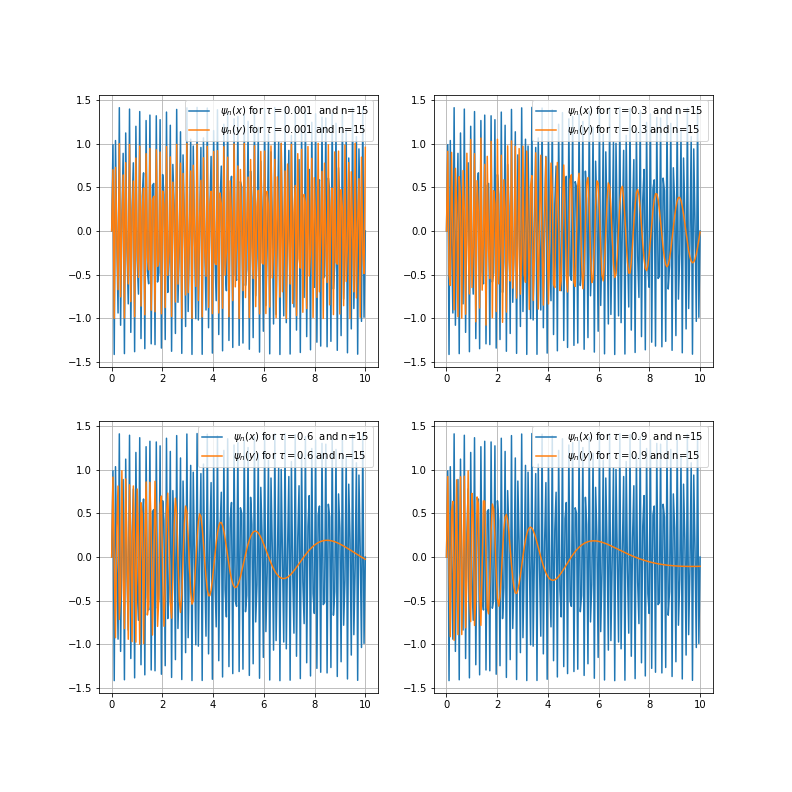}
  	 	  	 	}
  	 	  	 		\resizebox{0.5\textwidth}{!}{
  	 	  	 	  	 		\includegraphics{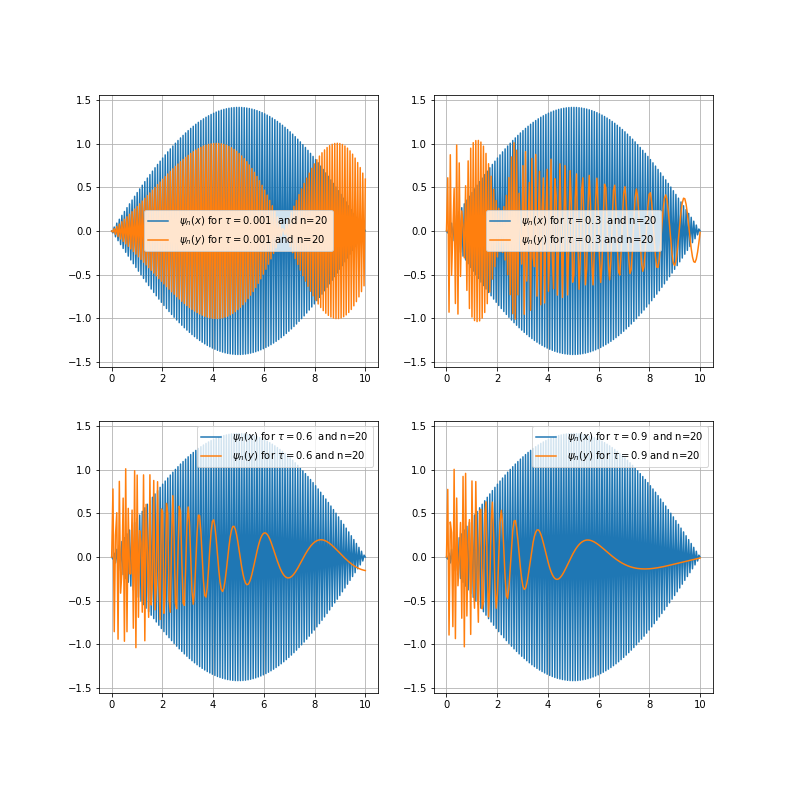}
  	 	  	 	  	 	}
  	 	  	 	  	 	
  	\caption{\it \small  Comparison graph of $\psi_n(x_s)$ and $\psi_n(y_s)$ for a particle confined in an infinite
  	    	square well of length $a = 1$ deformed by the gravity parameter $\tau$ in ys direction.
  	}
  	
  	\label{fig6}       
  \end{figure}

Figure \ref{fig6}  illustrates a comparison between eigenfunctions $\psi_n(x_s)$  and $\psi_n(y_s)$  for
fixed values of $n$ and $\tau$. The wave function in $x_s$-direction is taken as a witness with respect to that of the $y_s$-direction where the effects of quantum gravity are strongly applied. For $n\in \{1;5;15;20\}$, $\psi(y_s)$ compresses and contracts inward as
one increases $\tau$. This fact comes to confirm the fundamental property of gravity which is length contraction.

\begin{figure}[htbp]
    	\resizebox{0.5\textwidth}{!}{
    		\includegraphics{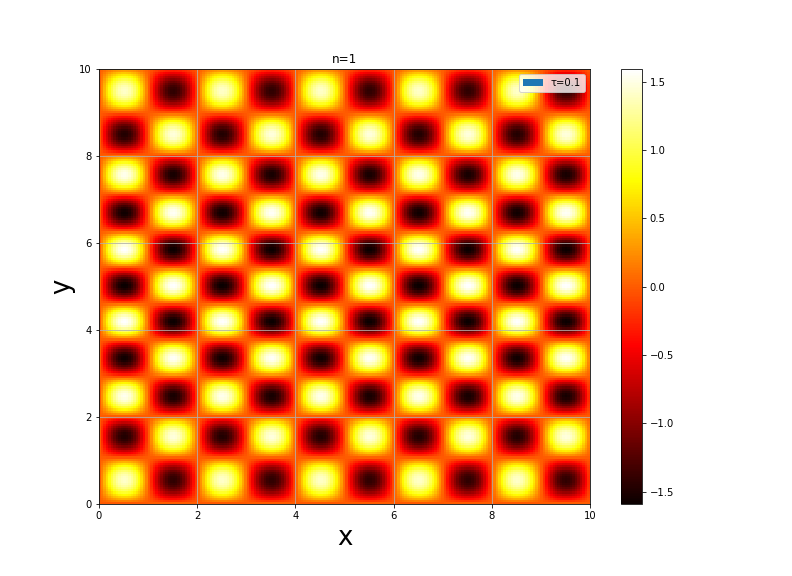}
    	}
    	\resizebox{0.5\textwidth}{!}{
    	 		\includegraphics{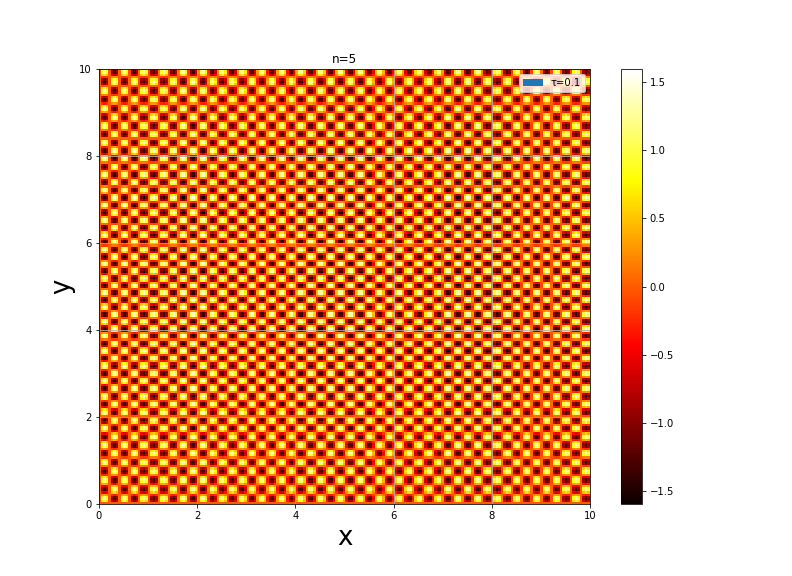}
    	 	}
    	 		\resizebox{0.5\textwidth}{!}{
    	 	  	 		\includegraphics{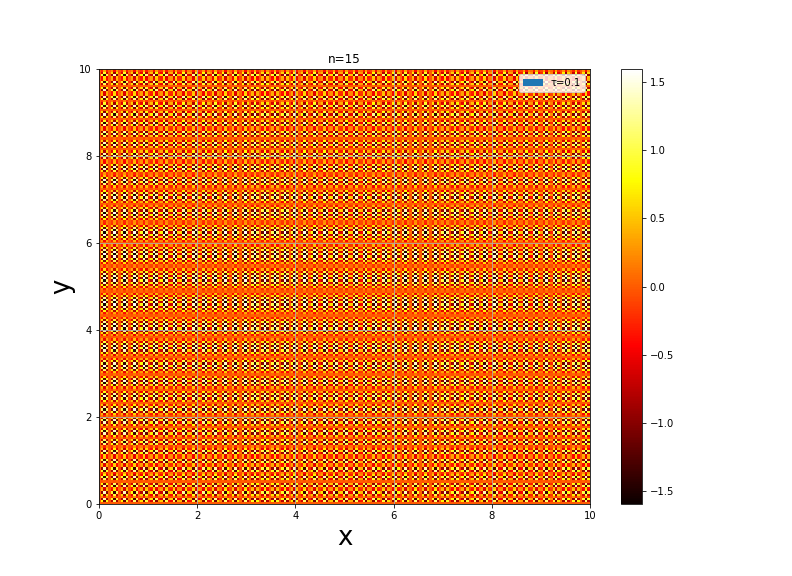}
    	 	  	 	}
    	 	  	 		\resizebox{0.5\textwidth}{!}{
    	 	  	 	  	 		\includegraphics{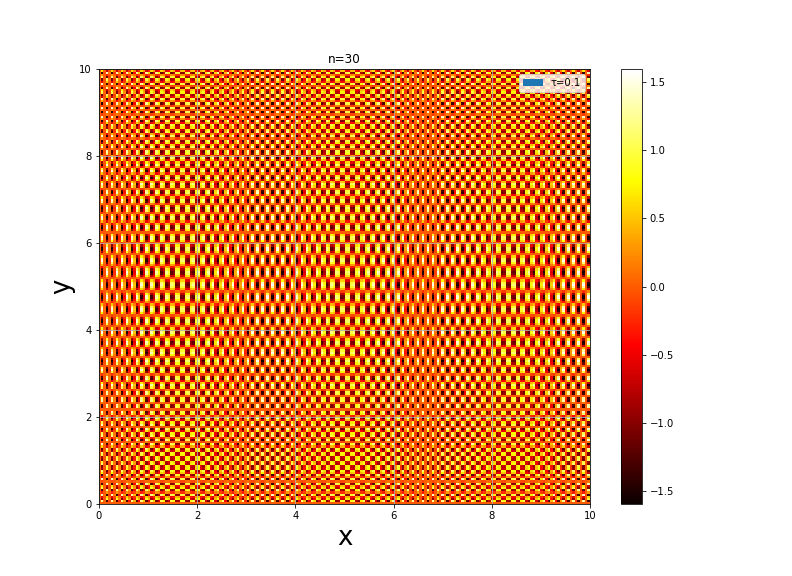}
    	 	  	 	  	 	}
    	 	  	 	  	 	
    	\caption{\it \small The probability density of a two-dimensional infinite square well for $\tau = 0.1$.
    	}
    	
    	\label{fig7}       
    \end{figure}

Unlike the figure reported in  \cite{57,58}, here the figure \ref{fig7} shows the plots of the
probability density for the three lower states $n = 1; n = 5; n = 15; n = 30$ for a
fixed value of $\tau$ ($\tau = 0.1)$, it can be seen that the probability to find a particle is practically the same everywhere in the square well and this probability strongly
increases with the quantum number. This indicates that the deformations allow  particles to jump from one state to another with low energies and with high
probability densities.
  
\section{Concluding remarks}
In this paper, we revisited our recent concept of minimal and maximal lengths
\cite{2}   (previously predicted by Perivolaropoulos \cite{1}) in the context of non-Hermitian position-dependent noncommutativity. We have shown that the existence of both
lengths has an interesting and significant properties of quantum gravity at the
Planck scale. We have  shown that a simultaneous measurement of both lengths form a lattice system in which each site represented by a minimal length $l_{min}$ is spaced by a maximal length $l_{max}$. At each singular point $l_{min}$ results
from the unification of strong magnetic  and weak quantum gravitational fields.
Furthemore, we have demonstrated that the maximal length of quantum gravity at this scale manifests properties similar to classical gravity and allows probing quantum gravitational effects with low energies.
Moreover the existence of minimal uncertainties 
lead almost unavoidably to non-Hermicity of some operators that generate the
noncommutative algebra (\ref{alg1}). Consequently, Hamiltonians of systems involving
these operators will not be Hermitian and the corresponding Hilbert space structure is modified. In order to map these operators into Hermitian ones, we have
introduced an appropriate Dyson map and by means of a similarity transformation, we have generated a hidden Hermitian position-dependent noncommutativity (\ref{alg3}). This transformation preserves properties of quantum
gravity at this  but  remove   the fuzziness
induced by the minimal uncertainties. 
Finally, to find the representation of a
free particle in this new space, we have solved a non-linear Schr\"{o}dinger equation.
To do so, we have transformed this equation into von Roos equation \cite{55}, then by
an appropriate change of variable, we reduced this equation into a simple and
solvable non-linear Schr\"{o}dinger equation.
 We observed that the increase of quantum  gravitational effects $\tau$ in this region curves
the quantum energy levels. These curvatures are more pronounced as one increases
the quantum levels, allowing the particle to jump from one state to another with
low energies and with high probability densities. Furthermore, they
 contract and compresse the wave function in  $y_s$-direction.
However, one can wonder about what happens in the case of a harmonic oscillator? In this way, the Hamiltonian of the system is given by
\begin{eqnarray} \label{Ham}
\hat H_{ho}=\frac{1}{2m_0}\left(\hat P_x^2+\hat P_y^2\right)+\frac{1}{2}m_0\omega^2(\hat X^2+\hat Y^2).
\end{eqnarray}
In terms of the $\theta$-deformed  variables, this  Hamiltonian can also be re-written as follows
\begin{eqnarray}
\hat H_{ho}&=&\frac{1}{2m_0}\left[\hat p_{x_0}^2+\left(1-\tau \hat y_0+\tau^2 \hat y_0^2\right)^2 \hat p_{y_0}^2-i\hbar \tau (1+2\tau \hat y_0)(1-\tau \hat y_0 +\tau^2\hat y_0^2)\hat p_{y_0} \right]+\cr&&\frac{1}{2}m_0\omega^2\left[\left(1-\tau \hat y_0+\tau^2 \hat y_0^2\right)^2 \hat x_0^2-i\theta (1-2\tau y_0)\left(1-\tau \hat y_0+\tau^2 \hat y_0^2\right)x_0+y_0^2\right].
\end{eqnarray}
 Since this Hamiltonian $\hat H_{ho}$ is  evidently non-Hermitian, we have to employ a Dyson map to convert it into Hermitian one   as in the previous example 
 
 \begin{equation}
 \hat h_{ho}=\eta \hat H_{ho}\eta^{-1}= \frac{1}{2m_0}\left(\hat p_x^2+\hat p_y^2\right)+\frac{1}{2}m_0\omega^2(\hat x^2+\hat y^2).
 \end{equation}
 This Hamiltonian may  be also  re-expressed as follows using the  representation (\ref{a1},\ref{a2},\ref{a3},\ref{a4})
 \begin{eqnarray}\label{Y}
 \hat h_{ho}&=&\frac{1}{2m_0}\left[\hat p_{x_0}^2+  (1-\tau\hat y_0 +\tau^2 \hat y_0^2)^{1/2}\hat p_{y_0} (1-\tau\hat y_0 +\tau^2 \hat y_0^2) \hat p_{y_0} (1-\tau\hat y_0 +\tau^2 \hat y_0^2)^{1/2}\right]+\cr&& \frac{1}{2}m_0\omega^2
 \left[\hat y_0^2+  (1-\tau\hat y_0 +\tau^2 \hat y_0^2)^{1/2}\hat x_0 (1-\tau\hat y_0 +\tau^2 \hat y_0^2) \hat x_0 (1-\tau\hat y_0 +\tau^2 \hat y_0^2)^{1/2}\right].
 \end{eqnarray}
The eigensystems of the equation (\ref{Y}) are far more complicated to obtain with
the same method as in the previous models as the system viewed as a differential equation no longer decouples in $x_0$ and $y_0$. We leave the construction of solutions
for this model by alternative means for future work.

\end{document}